\documentclass[conference]{IEEEtran}
  \usepackage[pdftex]{graphicx}
\usepackage{amsmath,amssymb,amsthm, latexsym}
\usepackage[font=small,labelfont=bf]{caption}
\usepackage{hyperref}

\usepackage[english]{babel}
\usepackage[utf8]{inputenc}
\usepackage{algorithm}
\usepackage[noend]{algpseudocode}
\usepackage{xcolor}
\usepackage[super]{nth}
\renewcommand\thesection{\Roman{section}} 
\renewcommand\thesubsectiondis{\thesection.\Alph{subsection}}
\renewcommand\thesubsubsectiondis{\thesubsectiondis.\arabic{subsubsection}}

\begin{document}
%
% paper title
% Titles are generally capitalized except for words such as a, an, and, as,
% at, but, by, for, in, nor, of, on, or, the, to and up, which are usually
% not capitalized unless they are the first or last word of the title.
% Linebreaks \\ can be used within to get better formatting as desired.
% Do not put math or special symbols in the title.
\title{An interleaver design for polar codes over slow fading channels}
% author names and affiliations
% use a multiple column layout for up to three different
% affiliations
\author{\IEEEauthorblockN{Saurabha R. Tavildar}
\IEEEauthorblockA{Email: tavildar at gmail }
}

\thispagestyle{plain}
\pagestyle{plain}
% make the title area
\maketitle
\begin{abstract}
~We consider the problem of using polar codes over slow fading wireless channels. For design, we focus on a parallel slow fading channel with $2$ blocks, and polar codes with rate $\leq \frac{1}{2}$. Motivated by Arikan's systematic polar code construction, we propose an interleaver design for a general polar code. The interleaver comprises of using the bit reversal of the order of polarized bit channels. This interleaver is called a diversity interleaver. In addition to the diversity interleaver, a diversity polar code is proposed to further increase the diversity gain.

The proposed designs are evaluated via link simulations for AWGN and fading channels. The simulation results show a performance close to the outage probability (within $\sim 2$ dB) and significant gains over using a random interleaver.
\end{abstract}

\section{Introduction}
% no \IEEEPARstart

Polar codes, introduced in \cite{Arikan}, were proved to achieve the symmetric capacity for BDMCs. Following \cite{Arikan}, works in \cite{MT}, \cite{Trifonov} have considered design of polar codes for AWGN channel. In this paper, we consider design of polar codes for fading channels. This has been considered in \cite{Santos}, \cite{Si} \cite{Trifonov2}. These works focus on polar code design for Rayleigh fading channels. The approach of designing codes targeted to a fading distribution can be challenging in context of a wireless system. This is because fading distribution varies dramatically based on many aspects such as deployment, carrier frequency, and mobility. Therefore, a universal design that is agnostic to the fading distribution is needed. 

Towards this goal, various works have considered maximizing diversity of a code focusing on maximizing performance when the channel is not in outage. In context of LDPC codes, a design to maximize code diversity was presented in \cite{Boutros-LDPC}. The work in \cite{Boutros-LDPC}, focuses on an LDPC code design that achieves full diversity. Our work has a similar objective as the work in \cite{Boutros-LDPC}, but for polar codes. In context of polar codes, \cite{Boutros-Polar} considers performance of polar codes over block fading channels. In other works, \cite{Wasserman} considers a variety of interleavers for polar codes for an OFDM system. We will compare our approach with approaches considered in \cite{Boutros-Polar} and \cite{Wasserman}. 

An overview of prior work is provided in Section~\ref{sec:prior}. Our proposed design is given in Section~\ref{sec:proposed}, and simulation results are given in Section~\ref{sec:sim}.

\section{Prior work}\label{sec:prior}

\subsection{Code design}

Tthe work in \cite{Boutros-LDPC} proposed a diversity achieving LDPC code for block fading channels for 2 blocks.  At a high level, the objective was to design a code that performs well even if one of the two blocks experiences a deep fade (the limiting case of "block-erasure channels"). This is the approach we take for polar code design as well. For polar codes, the work in \cite{Boutros-Polar} considered polar code design in the context of various interleaver choices. This is summarized next.

\subsection{Interleaver design}\label{sec:interleaver}

%Here, we summarize possible choices for interleaver design. In particular, we consider two related works in \cite{Boutros-Polar} and \cite{Wasserman}. In \cite{Boutros-Polar}, a two block fading model is considered. For this model, a choice of multiple multiplexers is considered and analyzed. In \cite{Wasserman}, a choice of inerleavers is considered for an OFDM system. In this paper, use the terminology of interleavers. 
%\begin{figure}[!ht]
%\includegraphics[width=\linewidth]{polar_interleaver_example.png}
%\caption{Possible interleavers (prior work)}\label{fig:prior_interleavers}
%\vspace*{0.05in}
%\end{figure}  

The following interleaver designs were considered in \cite{Boutros-Polar} and \cite{Wasserman} (with different terminology in some instances):
\begin{itemize}
\item None (called horizontal multiplexing in \cite{Boutros-Polar}): for this interleaver, the coded bits mapped in order. That is, the first $N/2$ coded bits are mapped to the first block, and the last $N/2$ coded bits are mapped to the second block. In our simulation results, this was the worst performing interleaver, and hence is not considered further in this paper;
\item Uniform (called diagonal multiplexing in \cite{Boutros-Polar}): for this interleaver, coded bits mapped uniformly. That is, all coded bits with an even valued index are mapped to the first block, and all coded bits an odd valued index are mapped to the second block;
\item Random (called uniform multiplexing in \cite{Boutros-Polar}): coded bits are pseudo-randomly interleaved; 
\item Block interleavers:  a sub-block interleaver for turbo codes for each component code in LTE \cite{212}. It is unclear if this class of interleavers is appropriate for polar codes. Simulation results in \cite{Wasserman} indicate that this is not a good choice for polar codes. We do not consider block interleavers further in this paper.
\end{itemize}

We note that a bit reversed mapping for coded bits as used in \cite{Arikan} is used in the description above and in the rest of this paper. We focus on uniform and random interleavers for comparison in this paper. %We note that for the work in \cite{Boutros-Polar} re-selected the information bits based on the choice of the interleaver.

\section{Proposed design}\label{sec:proposed}

We consider polar codes of rate  $\leq \frac{1}{2}$ and block-length $N$. For interleaver design, we focus on the block fading model of \cite{Boutros-LDPC}. In particular, we focus on the limiting case of "block-erasure channels" as discussed in \cite{Boutros-Polar}. For this case, the question of interleaver design reduces to selection of two subsets of code bits such that each of the subsets is adequate to decode the information bits for successive cancellation decoder (SCD). We call the subsets that satisfy this property as self decodable. 

Motivated by this property, we first propose an interleaver design, called diversity interleaver, for an arbitrary polar code in Section~\ref{sec:div_int}. In general, however, the two subsets of the diversity interleaver are not self decodable. To fix this, a construction of diversity polar codes is provided in Section~\ref{sec:div_polar}.

\subsection{Diversity interleaver}\label{sec:div_int}

Consider a polar code with rate exactly $\frac{1}{2}$. Let $A$ be the set of information bits for the polar code. Let $A^c$ be the complement of set A. For this code, one choice of a self decodable subset is based on the systematic polar code design of Arikan \cite{Arikan-Sys}. In particular, we note the following from \cite{Arikan-Sys}:

\vspace*{0.05in}

\noindent\fbox{%
    \parbox{\linewidth}{%
Another method for systematic encoding is to use a successive
cancellation decoder as an encoder. For this, one pretends
that x has been sent across a binary erasure channel (BEC)
and that the user data part $x_A$ has been received intact while
the remaining part $x_{A^c}$ has been fully erased. One initializes
the decoder suitably to reflect full knowledge of $(u_{A^c} , x_A)$
and complete uncertainty about $(u_A, x_{A^c})$. It can be shown
by recursive arguments that the decoder will always find $u_A$
correctly.
    }%
}
\vspace*{0.05in}

This suggests that $B(A)$ is self decodable where $B$ is the bit reversal mapping (note: \cite{Arikan-Sys} does not use the bit reversal mapping in the description above). This, however, only answers half of the question. For general polar codes, it is unclear if $B(A^c)$ is also self decodable. Nevertheless, we propose $B(A)$ and $B(A^c)$ to be the diversity interleaver. For polar codes with rate $< \frac{1}{2}$, we propose $A$ to be the highest $N/2$ reliability channels as per the code construction method. 

To summarize, the diversity interleaver is defined by:
\begin{itemize}
\item the set, $A$, of $N/2$ most reliable polarized bit channels;
\item the bit reversal mapping, $B$;
\item coded bits $B(A)$ sent over the first block;
\item coded bits $B(A^c)$ sent over the second block.
\end{itemize}

\subsection{Diversity polar codes}\label{sec:div_polar}

The main weakness of the diversity interleaver is that it is unclear if $B(A^c)$ is self decodable. To fix this problem, we propose diversity polar codes. In this construction, the polar code construction (for rate $\frac{1}{2}$) is limited by the following restriction on the set of information bits, $A$:
\begin{eqnarray}\label{eq:div_polar}
A^c & = & N + 1 - A.
\end{eqnarray}
That is, at most one of index $i$ or index $N + 1 - i$ is an information bit (index starting from $1$ as in \cite{Arikan} is assumed). This is done by picking the most reliable polarized bit channels while respecting constraint imposed by Equation~(\ref{eq:div_polar}). For rate $< \frac{1}{2}$, the set $A$ is chosen to be of size $N/2$ for the purposes of defining an interleaver. The set of information bits is then selected to be the most reliable bits within the set $A$. The diversity interleaver then is defined according to the modified set $A$ (instead of the most reliable $N/2$ channels).

One consequence of Equation~(\ref{eq:div_polar}) is that $B(A) = N + 1 - B(A^c)$. This makes sets $B(A)$ and $B(A^c)$ as equivalent as per Theorem 1 of \cite{Tavildar}, and suggests that both $B(A)$ and $B(A^c)$ are self decodable for diversity polar code. 

It is worth emphasizing that we do \textbf{not} provide a proof of self decodability of diversity interleaver for diversity polar code. We only conjecture that the two subsets, $B(A)$ and $B(A^c)$ are self decodable. It is unclear that under the limitation imposed by Equation~(\ref{eq:div_polar}), either $B(A)$  or $B(A^c)$ is self decodable.

\section{Simulation results}\label{sec:sim}
We simulate rate $\frac{1}{2}$ polar codes for block length values of $256$ and $1024$. Two polar code designs are considered:
\begin{itemize}
\item AWGN code: designed for BI-AWGN channel via density evolution. Design SNR chosen to target $\sim 1 \%$ BLER;
\item Diversity code: designed for BI-AWGN channel via density evolution while incorporating Equation~(\ref{eq:div_polar}).
\end{itemize}

Three interleaver designs are considered:
\begin{itemize}
\item Uniform interleaver as described in Section~\ref{sec:interleaver};
\item Random interleaver as described in Section~\ref{sec:interleaver};
\item Diversity interleaver as described in Section~\ref{sec:div_int}.
\end{itemize}

With exception of diversity code (as described in Section~\ref{sec:div_polar}), the polar codes are not re-designed based on the interleaver. Following three channel models are considered:
\begin{itemize}
\item AWGN;
\item Block Rayleigh: 2 i.i.d.\ blocks each of size $N/2$;
\item Frequency hopped OFDM (follows LTE numerology): 
 \begin{itemize}
      \item{2 RBs each 180KHz x 0.5ms separated by 10MHz;}
      \item{EPA-5 model with 1x2 antennas (0.5 correlation).}
    \end{itemize}
\end{itemize}
We note that simulating multiple channel models is important since in many systems, the same code is mapped differently depending on available resources. Further the same code will be used across a multiple deployment conditions. Therefore, it is important to have a good performance under multiple scenarios. 

Additional simulation assumptions are:
\begin{itemize}
\item BPSK modulation;
\item perfect CSI at the receiver;
\item successive cancellation decoder with no CRC.
\end{itemize}
In addition, outage probability is plotted for reference. The outage probability is defined as the probability that the mean instantaneous BPSK constellation capacity is less than the rate. 

\subsection{AWGN results}

\begin{figure}[!ht]
\begin{center}
\includegraphics[width=\linewidth]{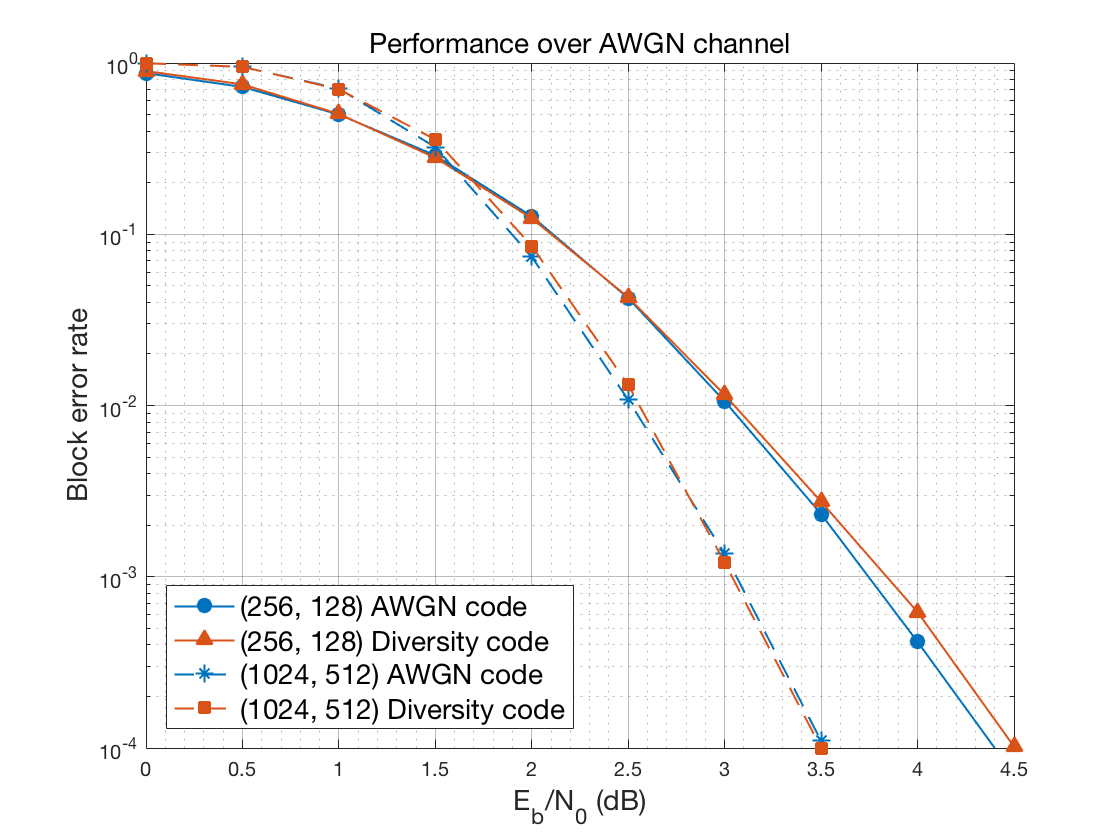}
\caption{AWGN performance for $N=1024, 256$, rate = $\frac{1}{2}$}\label{fig:diversity_1024_and_256_awgn}
\vspace*{0.05in}
\end{center}
\end{figure}

Figure~\ref{fig:diversity_1024_and_256_awgn} shows comparison for AWGN channel of the two code constructions. The main result is that diversity code performance is close to the AWGN code performance.  That is the limitation imposed by Equation~(\ref{eq:div_polar}) does not significantly impact performance for the AWGN channel.

\subsection{Block Rayleigh fading results}

\begin{figure}[!ht]
\begin{center}
\includegraphics[width=\linewidth]{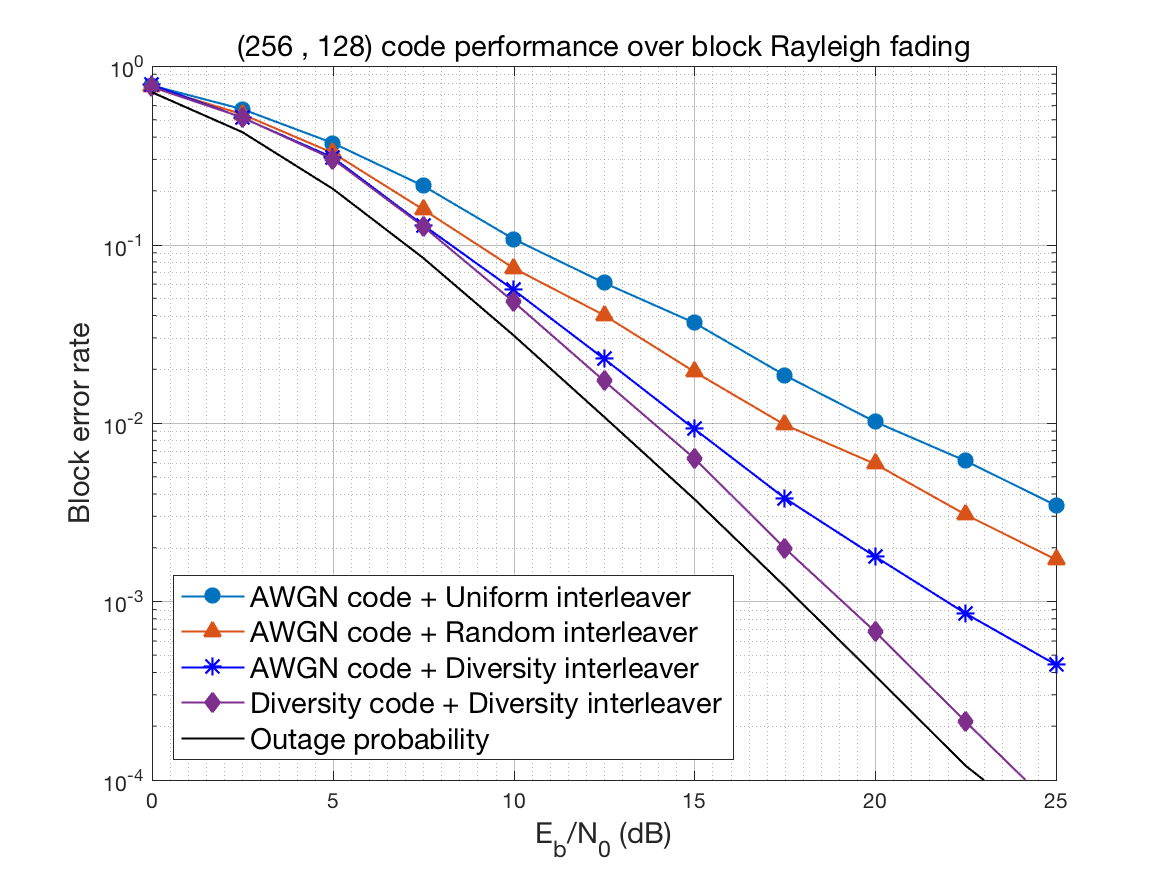}
\caption{Rayleigh performance for $N=256$, rate = $\frac{1}{2}$}\label{fig:diversity_256_fading}
\vspace*{0.05in}
\end{center}
\end{figure}

\begin{figure}[!ht]
\begin{center}
\includegraphics[width=\linewidth]{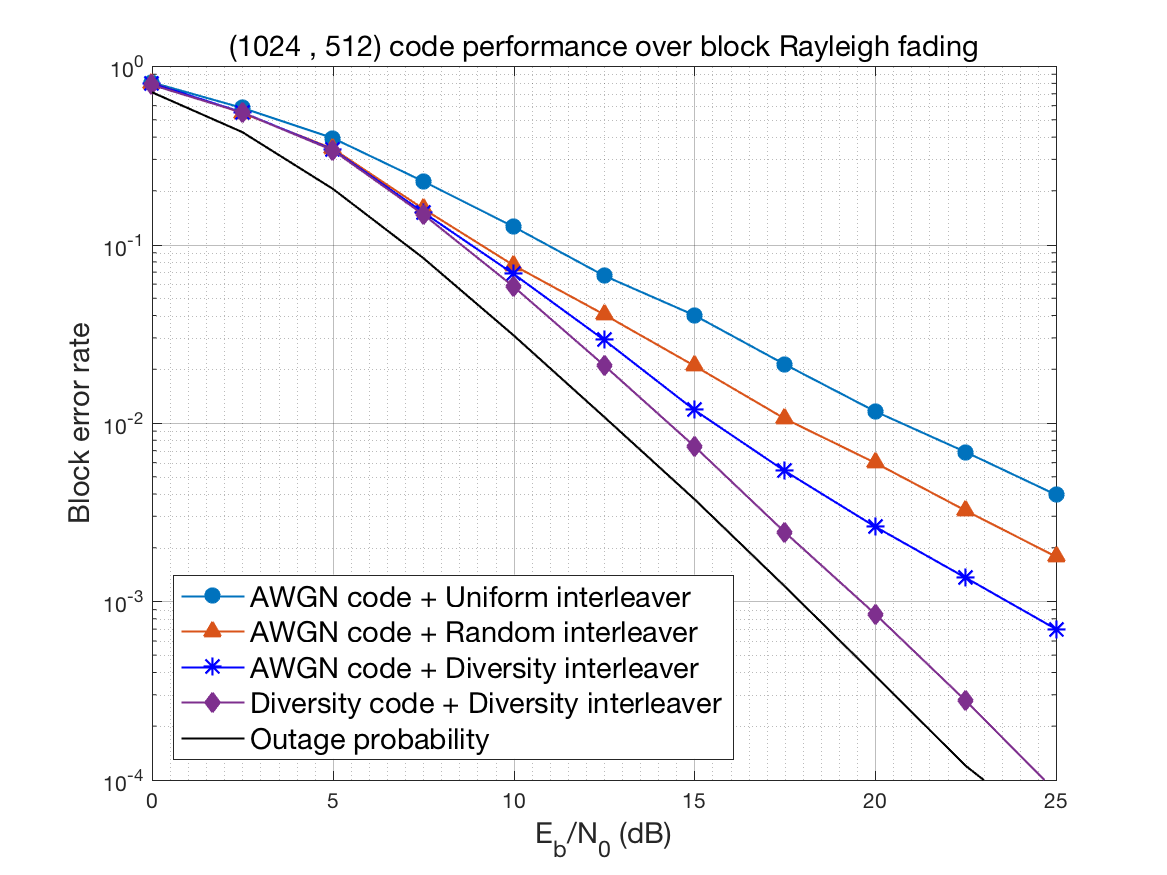}
\caption{Rayleigh performance for $N=1024$, rate = $\frac{1}{2}$}\label{fig:diversity_1024_fading}
\vspace*{0.05in}
\end{center}
\end{figure}

Figures~\ref{fig:diversity_256_fading} and \ref{fig:diversity_1024_fading} show comparison for block Rayleigh fading. The main results are:
\begin{itemize}
\item Diversity code with diversity interleaver has the best performance that is close to outage performance;
\item AWGN code with diversity interleaver has the next best performance;
\item AWGN code with random interleaver performs better than AWGN code with uniform interleaver.
\end{itemize}

\subsection{Frequency hopped OFDM}

\begin{figure}[!ht]
\begin{center}
\includegraphics[width=\linewidth]{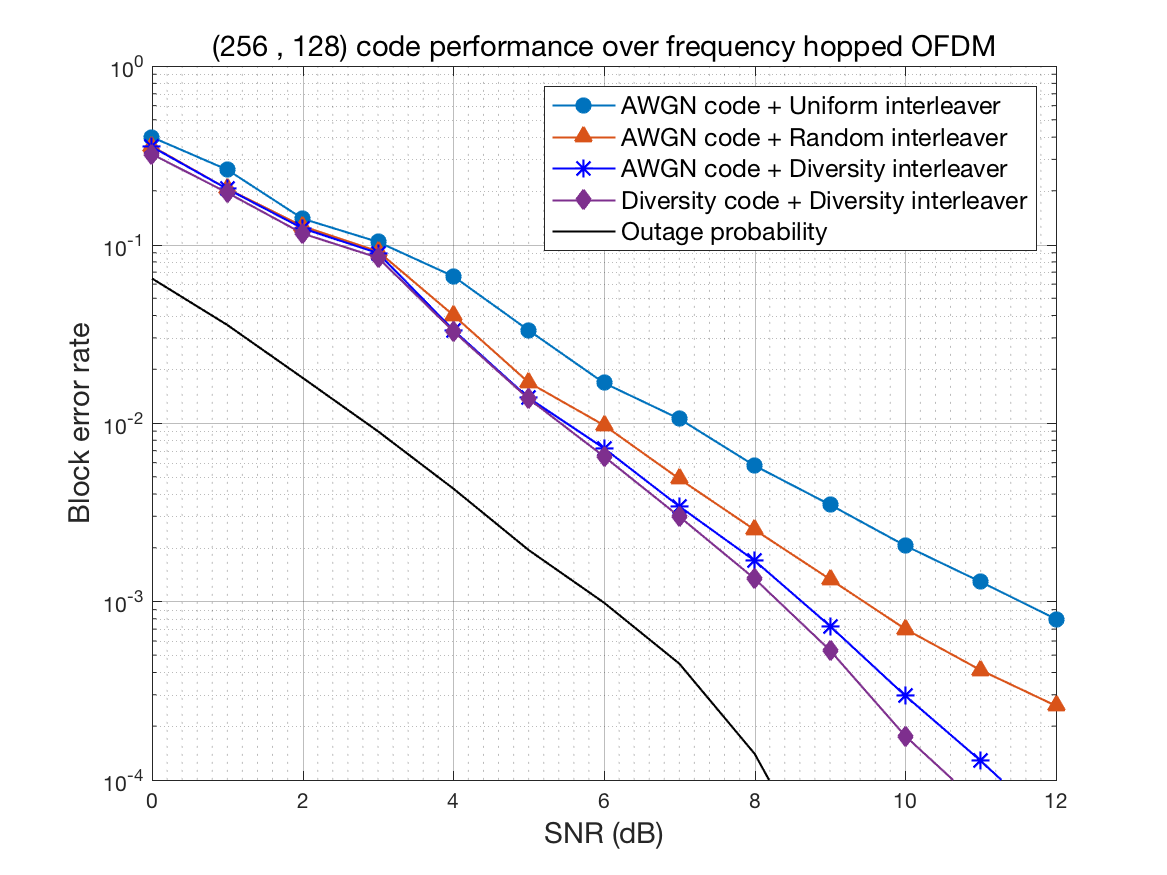}
\caption{Hopped OFDM performance for $N=256$, rate = $\frac{1}{2}$}\label{fig:diversity_256_128_ofdm}
\vspace*{0.05in}
\end{center}
\end{figure}

Finally, we simulate an OFDM system. We assume perfect equalization and perfect channel estimate. Only block length value of 256 is simulated. The results shown in Figure~\ref{fig:diversity_256_128_ofdm} are in-line with the results seen for block Rayleigh fading, but with smaller gains.

% that's all folks
\end{document}